\documentclass[%
 reprint,
superscriptaddress,
 amsmath,amssymb,
 aps,
 pra,
]{revtex4-1}

\usepackage{graphicx}% Include figure files
\usepackage{dcolumn}% Align table columns on decimal point
\usepackage{bm}% bold math
\usepackage{braket}
\usepackage{amsmath}
\usepackage{mathtools}
\usepackage{tikz}
\renewcommand{\d}{\mathrm{d}}
\newcommand{\e}{\mathrm{e}}
\renewcommand{\i}{\mathrm{i}}

\usepackage{subcaption}
\begin{document}

%\preprint{APS/123-QED}

\title{Hidden non-Markovianity in open quantum systems}

\author{Daniel Burgarth}%
\email{daniel.burgarth@mq.edu.au}
\affiliation{%
	Department of Physics and Astronomy, Center of Engineered Quantum Systems, Macquarie University,
	Sydney, NSW 2109, Australia
}%

\author{Paolo Facchi}
\email{paolo.facchi@ba.infn.it}
\affiliation{Dipartimento di Fisica and MECENAS, Universit\`{a} di Bari, I-70126 Bari, Italy}
\affiliation{	INFN, Sezione di Bari, I-70126 Bari, Italy}%

\author{Marilena Ligab\`o}%
\email{marilena.ligabo@uniba.it}
\affiliation{%
	Dipartimento di Matematica, Universit\`a di Bari, I-70125 Bari, Italy
}%

\author{Davide Lonigro}%
 \email{davide.lonigro@ba.infn.it}
\affiliation{Dipartimento di Fisica and MECENAS, Universit\`{a} di Bari, I-70126 Bari, Italy}
\affiliation{	INFN, Sezione di Bari, I-70126 Bari, Italy}%

\date{\today}% It is always \today, today,

\begin{abstract}
We show that non-Markovian open quantum systems can exhibit exact Markovian dynamics up to an arbitrarily long time; the non-Markovianity of such systems is thus perfectly ``hidden'', i.e. not experimentally detectable by looking at the reduced dynamics alone. This shows that non-Markovianity is physically undecidable and extremely counterintuitive, since its features can change at any time, without precursors. Some interesting examples are discussed.
%\begin{description}
%\end{description}
\end{abstract}

\pacs{Valid PACS appear here}% PACS, the Physics and Astronomy
\maketitle

%\tableofcontents

\section{Introduction}
The recent advance in quantum technology brought with it a renewed interest in the study of quantum noise. Never before have we built such complex high dimensional quantum systems, which naturally come with spatially and temporally correlated noise; and never before have we demanded such purity in quantum dynamics required for scalable quantum computation. Simplistic error models no longer suffice to achieve optimal performance~\cite{Modi1}.

A particular noise feature, the analysis of the Markovianity (or lack thereof) of the system, is of primary interest. A continuous process is said to be Markovian if, starting from any initial state, its evolution at any future time is determined unambiguously from the initial state, rather than by the full history of the system that led it to the present state. The lack of Markovianity is inherently linked with the two-way exchange of information between the system and the bath; a Markovian description is legitimate, even if only as an approximation, whenever the observed time scale of the evolution is much larger than the correlation time that characterizes the interaction between system and bath. Non-Markovianity is a complex phenomenon which affects the system both in its dynamical and informational features; several nonequivalent definitions of non-Markovianity, each focusing on particular aspects of memory, have been given. For a recent review we refer to~\cite{LiLi}. 

Non-Markovianity was discussed in a variety of physical systems and experimental platforms, such as cold atoms~\cite{Dorner02,zeta}, superconducting qubits~\cite{qdot_review,Tufarelli14}, photonic crystals~\cite{kimble1,ck}, waveguide quantum electrodynamics~\cite{waveguide,Fang18,n-wave}, optical fibers~\cite{Liu2018}, all-optical setups~\cite{Liu2011,Orieux2015,Bernardes2015,Cuevas19}, photonic waveguides~\cite{Zou2013,ladder},
the list being far from exhaustive. Most of these systems are well described by a paradigmatic theoretical model: 
the \emph{spin-boson model}, consisting of a two-level quantum system (qubit) interacting with a boson bath, the resulting rich phenomenology being ascribable to the structure of the bath and its interaction with the qubit.

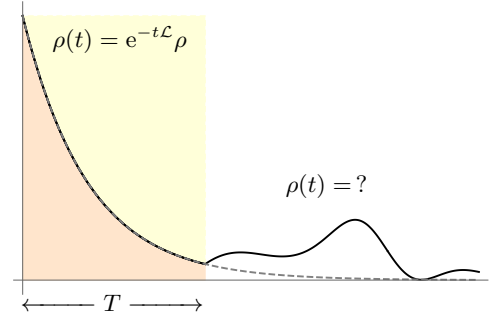
\begin{figure}[t]
	\begin{tikzpicture}
	\node at (0,0) {\includegraphics[scale=0.45]{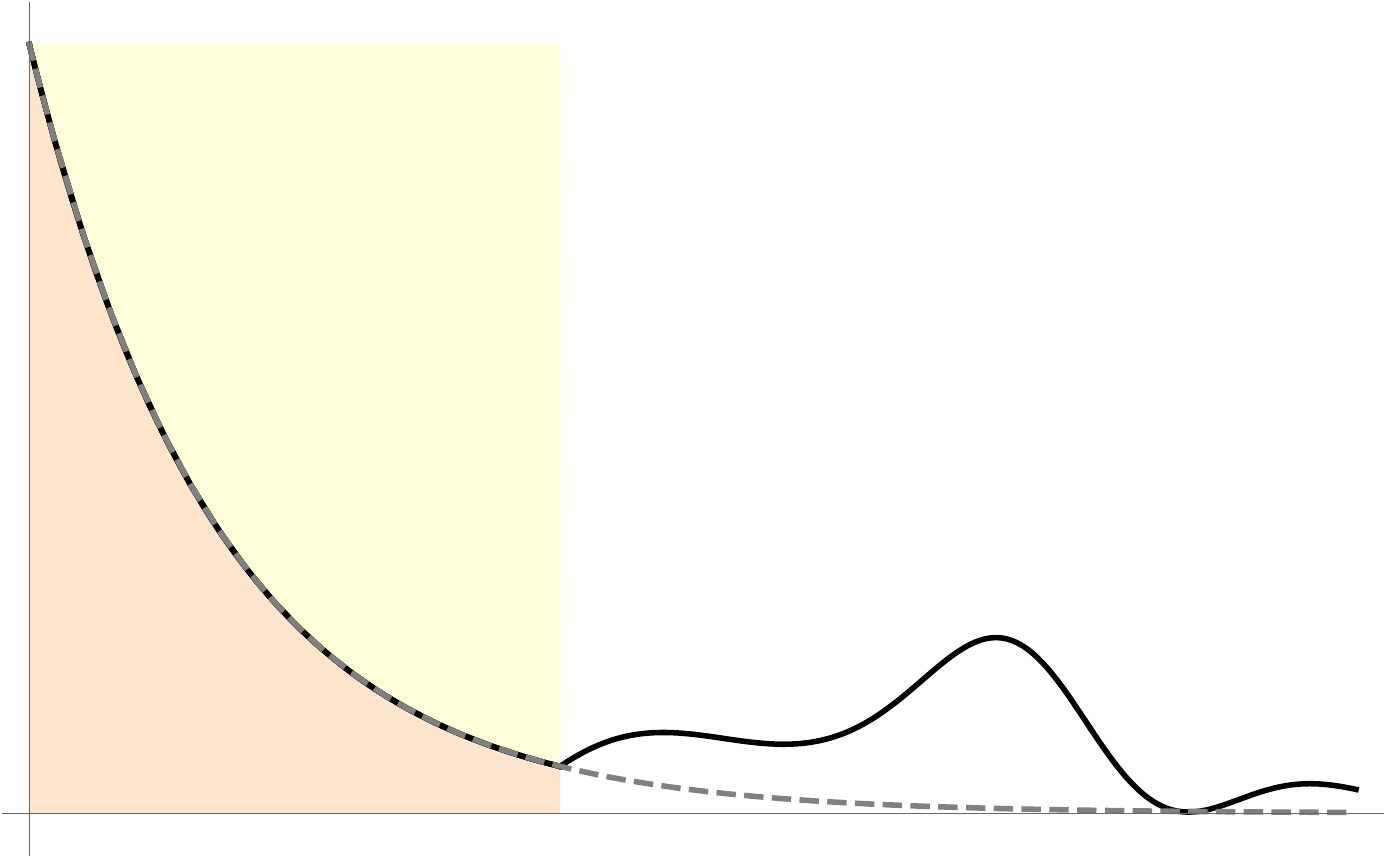}};
	\node at (-1.85,-2.05) {$\xleftarrow{\hspace*{0.8cm}} T\xrightarrow{\hspace*{0.8cm}}$};
	\node at (-1.75,1.45) {$\rho(t)=\e^{- t\mathcal{L}}\rho$};
	\node at (1.0,-0.5) {$\rho(t)=\,?$};
	\end{tikzpicture}
	\captionsetup{justification=raggedright,singlelinecheck=off}
	\caption{If a pure Markovian evolution is observed up to a time $t=T$, will the dynamics be Markovian for $t>T$ (dashed line) or might the dynamics deviate from Markovianity (solid line)?}
\end{figure}

Here we define a quantum evolution $\Lambda_t$ to be Markovian if it is described by a quantum dynamical semigroup,
$\Lambda_t=\e^{-t \mathcal{L}}$, with a time-independent generator $\mathcal{L}$~\cite{GKLS,Lindblad}. This narrow definition of Markovianity is a common core of many of the inequivalent definitions in the literature, although it is worth pointing out that such a definition does not capture the effect of time-dependent driving and other interventions~\cite{Modi}. The question if a fixed-time quantum operation $\Lambda_{t_0}$ (``snapshot'') can be embedded into a Markovian evolution  $\e^{-t_0 \mathcal{L}}$ was initiated in~\cite{Wolf}, and, remarkably, shown to be an NP-hard problem~\cite{Wolf2}. On the other hand, if more information, say the whole time \emph{evolution} $\Lambda_t$ for a time window $0\le t_1<t<t_2$, is provided, it \emph{appears} to be easy to decide Markovianity, simply by checking if the generator $-\Lambda_t^{-1}\frac{d}{dt}\Lambda_t$ exists and has time-independent Lindblad structure. The main point of our contribution  is to show that this is incorrect: deciding Markovianity remains hard for arbitrarily large windows. Without further knowledge on the environment or interventions on the dynamics it is, in fact, \emph{physically undecidable}. 

Recently, Tufarelli and co-authors~\cite{Tufarelli14} showed that there are systems which behave \emph{approximately} Markovian up to a critical time $T$, and non-Markovian thereafter. Although for time windows which do not exceed $T$ it is harder to assess non-Markovianity in such systems, they will still exhibit \emph{precursors} (in the spirit of~\cite{Precursor}) of non-Markovianity due to the coarse-graining of the Markovian approximation. That is, there will be slight deviations from the exact semigroup structure which reveal and anticipate the non-Markovianity at later time.

In this Article we will however show that, in fact, the spin-boson model can give rise to qubit evolutions which are \emph{exactly} Markovian up to some critical time $T$, without any precursor deviation of its dynamics. Since $T$ can be arbitrarily large, we conclude that   Markovianity of a quantum evolution \textit{cannot} be assessed, not even in the simplest case of a two-dimensional quantum system (qubit), by simply looking at the dynamics in a finite, however large, time window. In order to do so, we will construct explicitly a full family of non-Markovian quantum channels, for a qubit interacting with a given boson bath, whose dynamics is indistinguishable from the one induced by an exactly Markovian evolution up to a finite time. The reduced evolution of the qubit will be characterized by the following master equation:
\begin{equation}\label{eq:master}
\dot\rho(t) = -\i \varepsilon(t) \bigl[H_q, \rho(t) \bigr]  - \gamma(t) \mathcal{L} (\rho(t)),
\end{equation}
with $\rho(t)=\Lambda_t(\rho)$ being the density matrix of the qubit at time $t$, $\mathcal{L}$ being the  Lindblad super-operator~\cite{GKLS,Lindblad}  associated with an amplitude-damping channel, $H_{\mathrm{q}}$ the Hamiltonian of the qubit, and with $\gamma(t)$, $\varepsilon(t)$ being two real functions that only depend on the characteristics of the coupling between system and bath. 

In general, the quantum channel solving the master equation~\eqref{eq:master} will satisfy the semigroup property $\Lambda_{t+s}=\Lambda_t\Lambda_s$ at \textit{all} times $t,s\geq0$ only if the coupling is engineered in such a way that $\gamma(t)$ and $\varepsilon(t)$ are constant functions, which, as will be explained later, can only be obtained with an (essentially) unique choice of the coupling. However, there are \textit{infinitely} many ways to engineer the coupling in such a way that the semigroup property is satisfied \textit{only up to a finite time $T$}:
\begin{equation}\label{eq:hiddenmark}
	\Lambda_{t+s}=\Lambda_t\Lambda_s\qquad\text{for all }t,s\geq0, \quad t+s\leq T,
\end{equation}
with $T$ itself only depending on the choice of coupling. This can be obtained by choosing the coupling in such a way that the reduced dynamics of the system satisfies Eq.~\eqref{eq:master} with $\gamma(t)$, $\varepsilon(t)$ being constant only up to $t=T$. Such a system is by definition non-Markovian, but its non-Markovianity is \textit{hidden}: no observation at times $t\leq T$ will detect any deviation from Markovianity.

\section{The model}
We shall consider a qubit in a superposition of two orthogonal states $\ket{0}$ and $\ket{1}$, with $\omega_0$ being the energy gap between the two states, interacting with a boson quantum bath with creation and annihilation operators $b^\dagger_\omega$, $b_\omega$ satisfying the  commutation relations $[b_\omega,b_{\omega'}^\dagger]=\delta(\omega-\omega')$. The Hamiltonian has the form $H=H_0 + H_{\mathrm{int}}$, where
\begin{equation}
H_0 = \omega_0 H_{\mathrm{q}} \otimes \openone + \openone \otimes H_{\mathrm{B}}, 
\end{equation}
is the free Hamiltonian describing the uncoupled evolution of the qubit,
with $H_{\mathrm{q}} =  \ket{0}\!\bra{0}$, and the field, with $H_{\mathrm{B}} = \int \d\omega\, \omega\, b^\dag_{\omega}b_{\omega}$. The qubit-field interaction  has the form 
\begin{equation}
	H_{\mathrm{int}}= \sigma_+ \otimes B(g) + \sigma_- \otimes B(g)^\dagger, \quad
	B(g)=  \int\mathrm{d}\omega\, g(\omega)^* b_\omega,
\label{eq:Hint}
\end{equation}
where $\sigma_+=\ket{0}\!\bra{1}$, $\sigma_-=\ket{1}\!\bra{0}$, and the function $g(\omega)$, the \textit{form factor}, weights the strength of the interaction of the qubit with a boson of energy $\omega$. The interaction~\eqref{eq:Hint} has a rotating-wave form:  a boson with wavefunction $\ket{g(\omega)}$ is created if the qubit undergoes the transition $\ket{0}\rightarrow\ket{1}$, and is annihilated if the qubit undergoes the inverse transition $\ket{1}\rightarrow\ket{0}$. As a consequence the excitation number $N =  \sigma_{+} \sigma_{-} + \int \d\omega\, b^\dag_{\omega} b_{\omega}$ is conserved, $[N,H]=0$, so that the sectors with given excitation number are invariant under the evolution. In particular,  the component of the Hamiltonian in the one-excitation sector, known as the Friedrichs-Lee model, has very rich mathematical properties that have been extensively studied in~\cite{FriedLee,FriedLeeProc}. 

We will focus on the reduced dynamics induced by this Hamiltonian on a state $\rho\otimes\ket{\mathrm{vac}}\!\bra{\mathrm{vac}}$ by tracing out the bath, with the vector $\ket{\mathrm{vac}}$ being the vacuum of the boson field characterized by $b_\omega\ket{\mathrm{vac}}=0$ for all $\omega$. Define
\begin{equation}
\rho(t)= \Lambda_t (\rho) = \operatorname{tr}_{\mathrm{bath}}\left(\e^{-\i tH}\rho\otimes\ket{\mathrm{vac}}\!\bra{\mathrm{vac}}\e^{\i tH}\right);
\end{equation}
in the Appendix we show that $\rho(t)$ is given by~\cite{unbounded}
\begin{equation}
	\rho(t)=\begin{pmatrix}
	|a(t)|^2\rho_{00}& a(t)  \rho_{01}\\
	a(t)^*\rho_{10}&\rho_{11}+(1-|a(t)|^2)\rho_{00}
	\end{pmatrix},
\end{equation}
where $a(t)$ is a complex function with $a(0)=1$, $|a(t)|\leq1$ that is solely determined by the coupling function $|g(\omega)|^2$ and the energy of the state $\ket{0}$. 
Physically, 
$a(t)$ is the \textit{survival amplitude} of the state $\ket{0}$.

The density matrix $\rho(t)$ satisfies the master equation~\eqref{eq:master} with $\mathcal{L}(\rho) =  - \sigma_- \rho \sigma_+ + \frac{1}{2} \bigl\{ \sigma_+ \sigma_-, \rho \bigr\}$, $H_q = \ket{0}\!\bra{0}$, and the functions $\gamma(t)$ and $\varepsilon(t)$ being defined via
\begin{equation}
	a(t) = \e^{-\int_0^t \d s\, \left(\frac{\gamma(s) }{2}  + \i \varepsilon(s) \right) }.
\end{equation}
A particularly important case is a flat form factor, i.e. $|g(\omega)|^2=\gamma_0/2\pi$ for some $\gamma_0>0$: the qubit couples with the same strength to all frequencies of the boson field. Although the Hamiltonian is singular in such a case, one can prove that it yields a \textit{bona fide} unitary evolution  of the total system ($H$ is self-adjoint)~\cite{FriedLeeProc}. In fact, the qubit density matrix $\rho(t)$ satisfies Eq.~\eqref{eq:master} with $\gamma(t)=\gamma_0$ and $\varepsilon(t)=\varepsilon_0$ both being constant, where $\varepsilon_0 = \omega_0 + \delta\omega_0$ is the dressed energy of state~$\ket{0}$ in interaction with by the boson field~\cite{FriedLeeProc} (see Appendix), thus
\begin{equation}
	a(t)=\e^{-\left(\frac{\gamma_0}{2} + \i\varepsilon_0\right)t},
\end{equation}
and, in particular, the channel satisfies the semigroup property at \textit{all} times $t,s\geq0$, i.e. it is Markovian. 

However, we can choose the coupling $g(\omega)$ in such a way that $a(t)$ is exactly exponential \textit{only up to a finite time} $T$, and is no longer exponential afterwards; in such a way, Eq.~\eqref{eq:hiddenmark} holds and we obtain a non-Markovian system whose non-Markovianity is however \textit{hidden}: no experiment performed within the time horizon $T$ will be able to detect any deviation from the exponential law. This can be accomplished by choosing a \textit{periodic} coupling $|g(\omega)|^2$, whose Fourier series reads
\begin{equation}\label{eq:periodic}
	|g(\omega)|^2=\frac{\gamma_0}{2\pi}\biggl(1+2\sum_{n=1}^\infty c_n\cos (nT\omega) \biggr),
\end{equation}
where the Fourier coefficients are chosen in such a way that $0\leq|g(\omega)|^2<\infty$: the damping function $a(t)$ corresponding to this coupling is evaluated in the Appendix and reads
\begin{eqnarray}
\label{eq:solution}
a(t)&=&\e^{-\left(\i\varepsilon_0+\frac{\gamma_0}{2}\right)t}+\nonumber\\&+&\sum_{n=1}^\infty \e^{-\left(\i\varepsilon_0+\frac{\gamma_0}{2}\right)(t-nT)}\phi_n\bigl(\gamma_0(t-nT)\bigr) \theta(t-nT),\nonumber\\
\end{eqnarray}
where $\theta$ is the Heaviside step function; this function is exactly exponential up to $t=T$, while non-exponential terms start adding up at times $nT$, for $n=1,2,\dots$ In detail, here $\phi_n(x)$ is a polynomial of degree $n$ whose coefficient can be analytically computed in terms of the coefficients $c_n$.

Physically, the above behavior is a consequence of the time-energy uncertainty relation $\Delta t\, \Delta \omega \geq 1/2$ (a general property of the Fourier transform). Any measurement that lasts less than a time $T$ cannot resolve energy differences $\Delta\omega$ below $1/ (2 T)$. Therefore, the observation of the decay in a time window of width $T$ will depend on a \textit{coarse-graining} of the form factor. A coarse-grained periodic coupling will be indistinguishable from a flat one if the resolution is larger than its periodicity. Only for times larger than $T$ the system will start to resolve the finer details of a non-flat coupling and the underlying non-Markovianity will start to become manifest.

In the following we will furnish two explicit examples of form factors $g(\omega)$ for which all terms in~\eqref{eq:solution} can be evaluated explicitly.

\section{Two examples}
\begin{figure}[ht]
	\centering
	\begin{tikzpicture}
	\node at (0,0) {\includegraphics[scale=0.38]{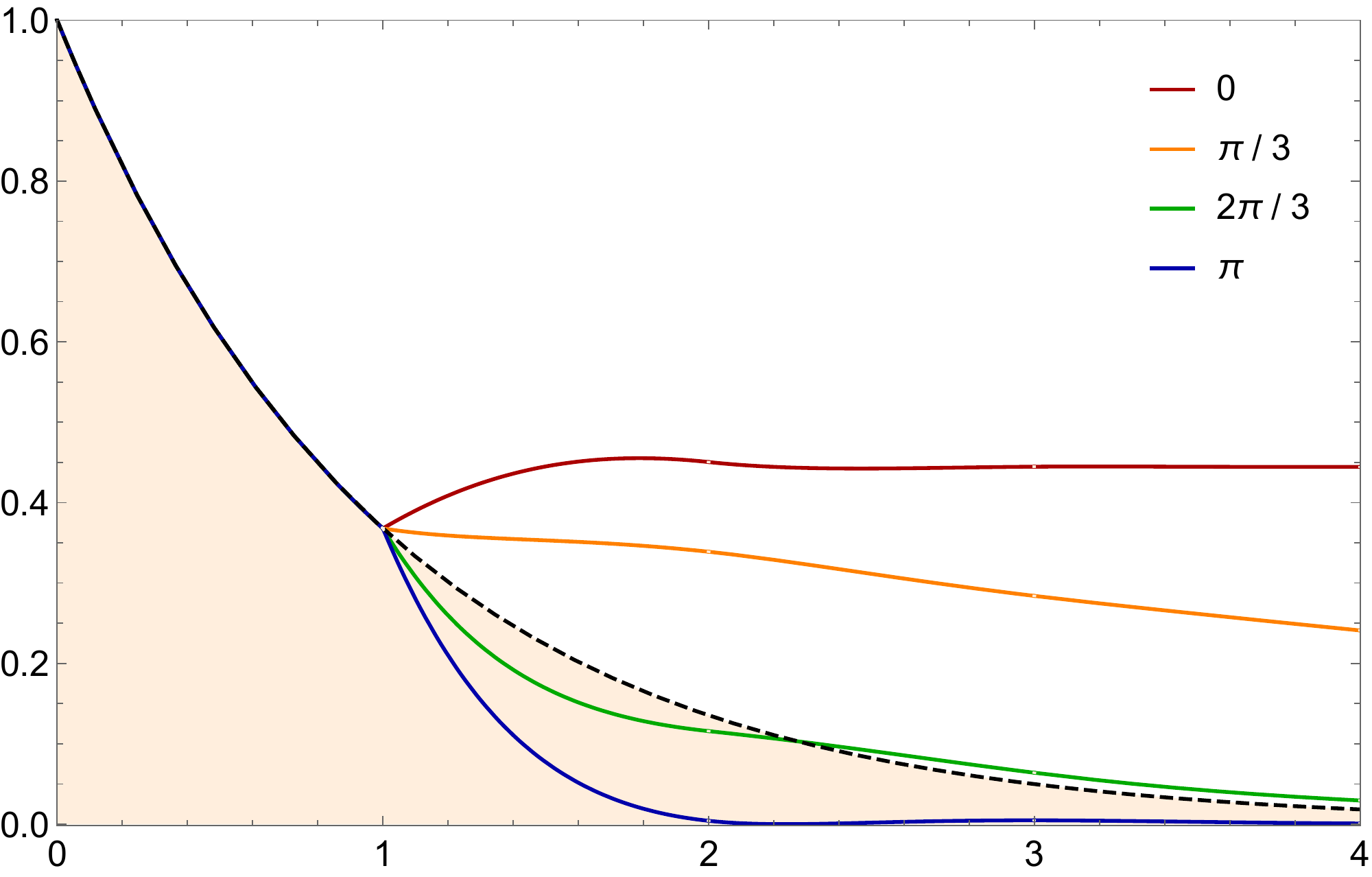}};
	\node at (0,-2.8) {$t/T$};
	\node at (-4.3,0) [rotate=90] {$|a(t)|^2$};
	\node at (1.3,1.95) {$\gamma_0T=1,\;\varepsilon_0T=$};
	\end{tikzpicture}
	\begin{tikzpicture}
	\node at (0,0) {\includegraphics[scale=0.38]{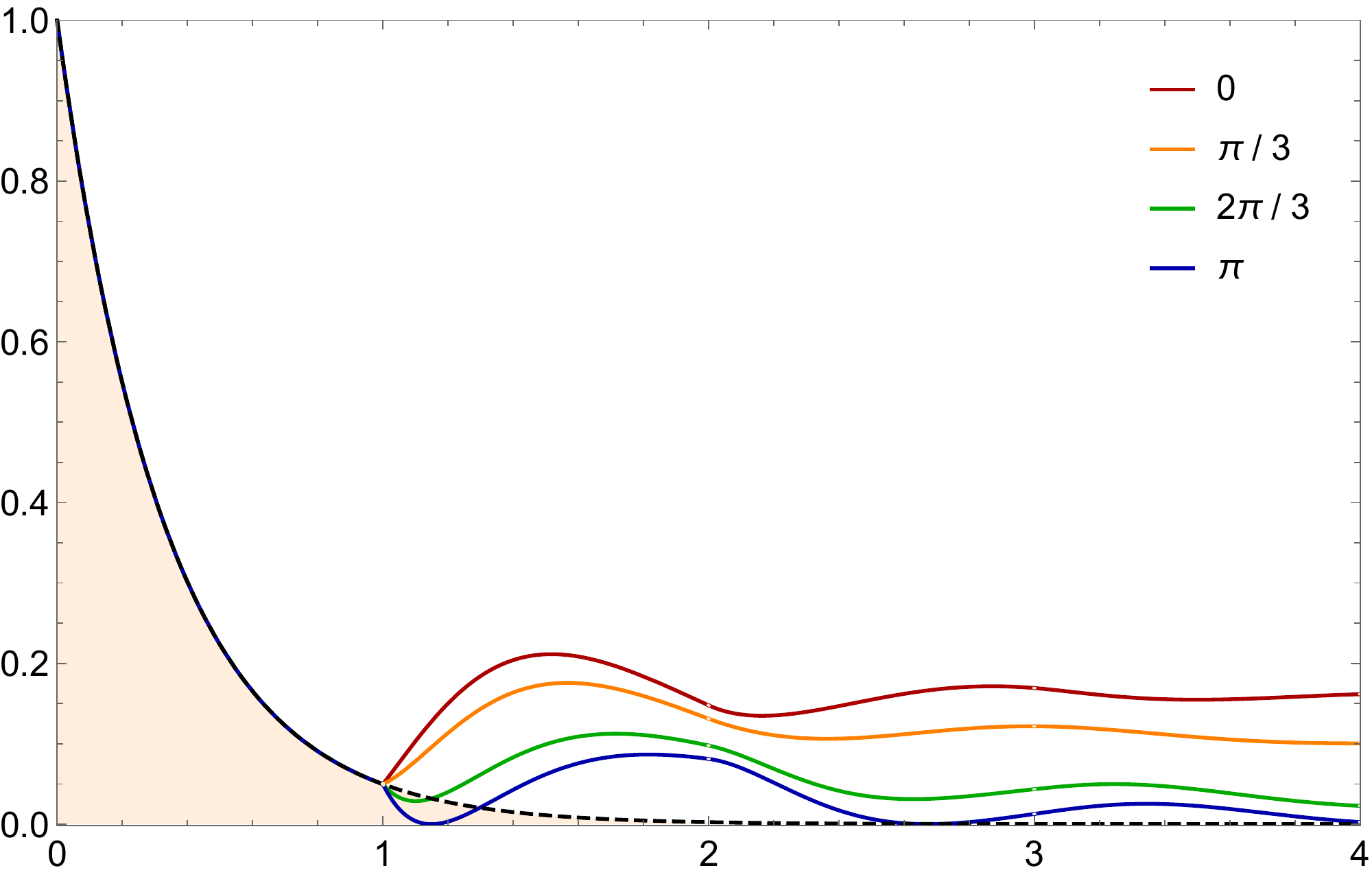}};
	\node at (0,-2.8) {$t/T$};
	\node at (-4.3,0) [rotate=90] {$|a(t)|^2$};
	\node at (1.3,1.95) {$\gamma_0T=4,\;\varepsilon_0T=$};
	\end{tikzpicture}
		\captionsetup{justification=raggedright,singlelinecheck=off}
	\caption{Survival amplitude $a(t)$ corresponding to a periodic coupling with Fourier coefficients as given in Eq.~\eqref{eq:sinusoidal}, with $\alpha=1$ and $\varepsilon_0 T = 0, \pi/3, 2\pi/3, \pi$ (mod $2\pi$).
	\label{fig:plotsin}}
\end{figure}
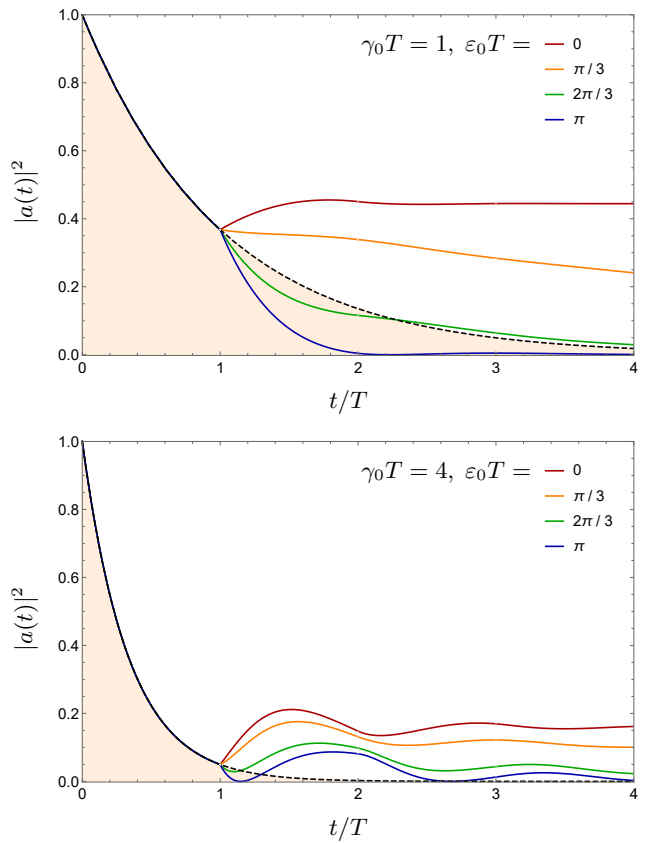
The simplest nontrivial example can be obtained by setting, in Eq.~\eqref{eq:periodic},
\begin{equation}\label{eq:sinusoidal}
	c_1=-\frac{\alpha}{2},\qquad c_{n}=0\text{ for all }n\geq2
\end{equation}
for some $|\alpha|\leq1$; in this case, $|g(\omega)|^2$ is a sinusoidal function whose amplitude is maximal for $\alpha=\pm1$ and null for $\alpha=0$. Physically, the choice $\alpha=1$ can be associated with a quantum emitter coupled with a semi-infinite waveguide with a perfect mirror at on end; the parameter $T$ will correspond to the time after which an emitted photon, reflected by the mirror, will reach the emitter again; the \textit{delay differential equation} (DDE) corresponding to the system was first obtained through some approximations in~~\cite{Dorner02}, while the non-Markovianity of the system was thoroughly investigated in~\cite{Tufarelli14,Fang18} via non-Markovianity measures. The case $\alpha=0$ corresponds again to a flat coupling, and thus to a Markovian evolution.

All polynomials $\phi_n(x)$ in Eq.~\eqref{eq:solution} have the simple form
$\phi_n(x)=\frac{1}{n!}\left(\frac{\alpha x}{2}\right)^n$,
(see Appendix) and thus the function $a(t)$ can be evaluated at all times; see Fig.~\ref{fig:plotsin}. The results can be summarized as follows. With respect to the pure exponential decay at $\alpha=0$, the decay will be either enhanced or slowed down depending on the values of the parameters $\alpha$, $\varepsilon_0$ and $\gamma_0$, and, in particular, for any fixed $\alpha$ the decay will be slowest when $g(\varepsilon_0)$ is smallest, i.e. when $\varepsilon_0T=2\nu\pi$ for some integer $\nu$. In particular, if $\alpha=1$ \textit{and} $\varepsilon_0=2\nu\pi$, $a(t)$ does not decay at all: a bound state is obtained. In the physical implementation of the model in waveguide QED, the emitter is at a distance of an integer number of half-wavelengths from the mirror and the photon is trapped between emitter and mirror. The departure from Markovianity is thus maximal.

\begin{figure}[ht]
	\centering
	\begin{tikzpicture}
	\node at (0,0) {\includegraphics[scale=0.38]{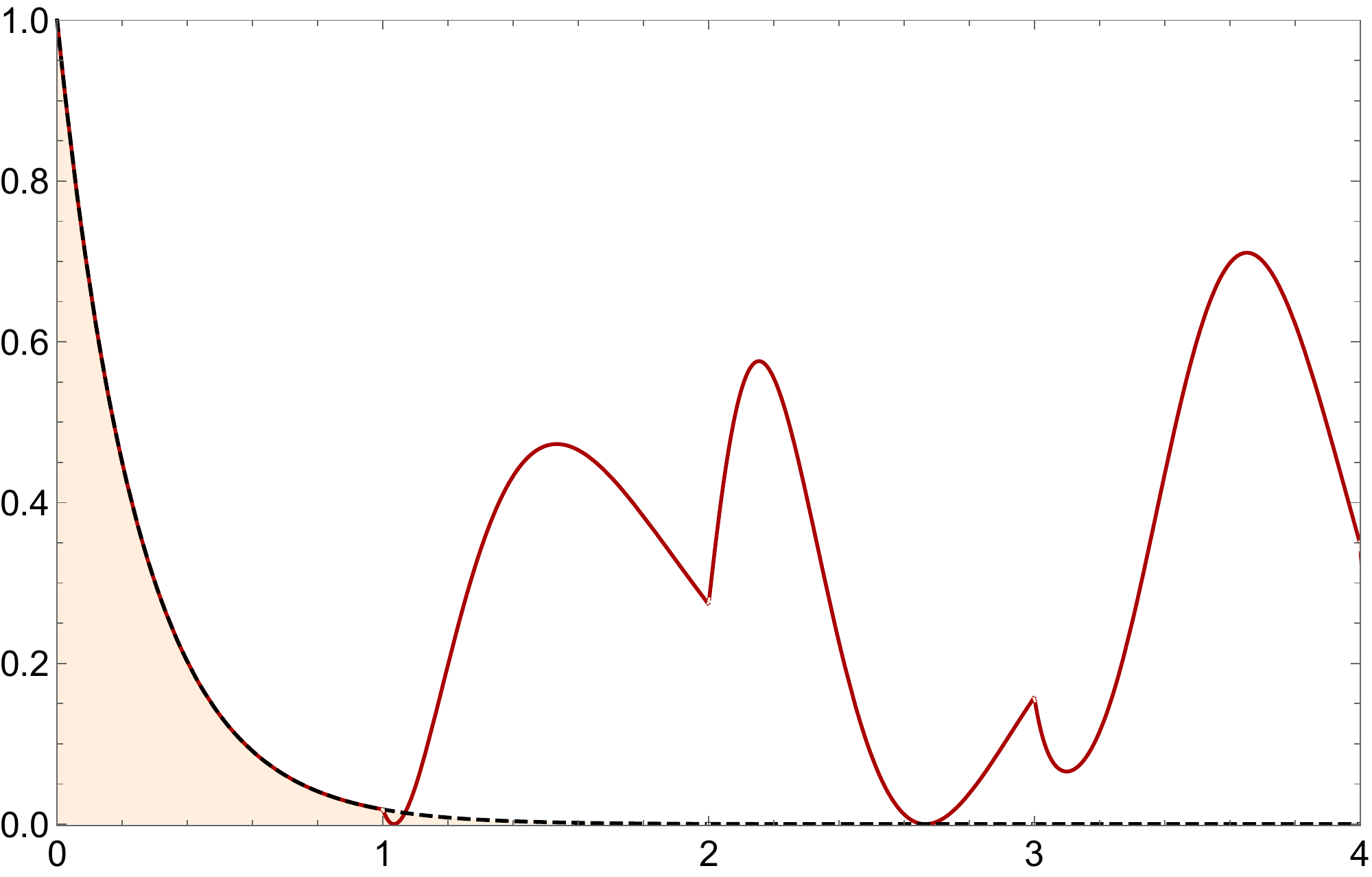}};
	\node at (0,-2.8) {$t/T$};
	\node at (-4.3,0) [rotate=90] {$|a(t)|^2$};
	\end{tikzpicture}
	\captionsetup{justification=raggedright,singlelinecheck=off}
	\caption{Survival amplitude $a(t)$ corresponding to a periodic coupling with Fourier coefficients as given in Eq.~\eqref{eq:smoothed}, with $\beta=0$, $\varepsilon_0T=0$ and $\gamma_0T=4$.}
\end{figure}
Another instance of periodic coupling for which $a(t)$ can be computed exactly is obtained by setting, in Eq.~\eqref{eq:periodic},
\begin{equation}\label{eq:smoothed}
	c_n=\e^{-\beta n}
\end{equation}
for some $\beta\geq0$. If $\beta=0$, this is a comb of Dirac functions placed at integer values of the energy, while for $\beta>0$ it is a ``smoothed'' comb. A physical implementation of the discrete case $\beta=0$ can be obtained by considering a closed loop waveguide or a one-dimensional optical cavity: indeed, when confining the boson field in a finite space, the emitter will only interact with a countable set of boson states. Interestingly enough, the DDE  for $\beta=0$ was already obtained in 1984 by Milonni and co-authors~\cite{Milonni83} in a different framework, and has been rediscovered afterwards a couple of times. The flat coupling is recovered in the opposite limit $\beta\to+\infty$. 

Again, with this class of couplings the dynamics is exactly computable at all times (see Appendix): Eq.~\eqref{eq:solution} holds with
	$\phi_n(x)=\e^{-\beta n}\sum_{m=1}^n\binom{n-1}{m-1}\frac{(-x)^m}{m!}$,
	implying that the non-Markovian contributions to the survival amplitude have the same functional expression for all $\beta$, up to a total weight $\e^{-\beta n}$ which suppresses such contributions as $n$ grows, provided that $\beta>0$; as a result, the larger $\beta$, the quicker such contributions ``switch off'', whereas for small $\beta$ those contributions are non-negligible for a longer time. In particular, in the limit $\beta\to\infty$ all non-Markovian contributions vanish and we recover the exponential decay at all times. In the opposite limit $\beta\to0$, where the coupling is discrete, no exponential suppression of such contribution happens and we have recurring dynamics with revivals at all times.
	
\section{Conclusions}
In this Article we show that no finite-time measurement can establish Markovianity of an open quantum system: the non-Markovianity may indeed be hidden, in the sense that non-Markovian effects may only switch on after some time threshold. To show this, we have considered a model of interaction between a qubit and a boson bath which reduces to an amplitude-damping channel for the former, with a survival amplitude which can be tuned by properly choosing the form factor of the coupling; whenever the latter is a periodic function, non-Markovian effects will only arise after a finite time. Remarkably, such corrections can be computed exactly: two particular examples have been discussed.

\begin{acknowledgments}
DB acknowledges support by the Australian Research Council (project number FT190100106). PF, ML and DL were partially supported by the Italian National Group of Mathematical Physics (GNFM-INdAM). PF and DL were partially supported by Istituto  Nazionale di Fisica Nucleare (INFN) through the project ``QUANTUM'', and by Regione Puglia and QuantERA ERA-NET Cofund in Quantum Technologies (GA No. 731473), project PACE-IN.
\end{acknowledgments}

\section*{Appendix}
\subsection{The model}
We shall consider a qubit, in a superposition of two orthogonal states $\ket{0}$ and $\ket{1}$, interacting 
with a bosonic quantum bath at zero temperature.
The microscopic Hamiltonian is $H = H_0+ H_{\mathrm{int}}$
where 
\begin{equation}
H_0= \omega_0 H_{\mathrm{q}} \otimes \openone + \openone \otimes H_{\mathrm{B}}
\end{equation}
and
\begin{equation}
H_{\mathrm{q}} = \sigma_{+}\sigma_{-} =  \ket{0}\!\bra{0}, \qquad H_{\mathrm{B}} = \int \d\omega\, \omega\, b^\dag_{\omega}b_{\omega},
\end{equation}
are the qubit Hamiltonian and the bath Hamiltonian, respectively, while
\begin{equation}
H_{\mathrm{int}} = \sigma_+ \otimes B(g) + \sigma_{-} \otimes B^\dag(g) 
\end{equation}
is the interaction Hamiltonian with
\begin{equation}
\sigma_+ = \sigma_{-}^\dag = \ket{0}\!\bra{1},\qquad B(g) = \int \d\omega\, g(\omega)^* b_{\omega},
\end{equation}
and $b_{\omega}$, $b^\dag_{\omega}$ are the bosonic annihilation and creation operators, satisfying the CCR $[b_{\omega}, b^\dag_{\omega'}]= \delta(\omega-\omega')$,  $[b_{\omega}, b_{\omega'}]=0$.
That is
\begin{eqnarray}\label{eq:ham}
H&=&\omega_0\ket{0}\!\bra{0}\otimes\openone+\openone\otimes\int   \d\omega\, \omega\, b^\dag_{\omega}b_{\omega}\nonumber\\& & + \ket{0}\!\bra{1}\otimes \int \d\omega\, g(\omega)^* b_{\omega} +  \ket{1}\!\bra{0}\otimes 
\int \d\omega\, g(\omega) b^\dag_{\omega}.\nonumber\\
\end{eqnarray}
Here $g(\omega)$ is a complex function that weights the strength of the interaction; the interaction term is constructed in such a way that a boson is created if the qubit undergoes the transition $\ket{0}\rightarrow\ket{1}$, and is annihilated if the qubit undergoes the transition $\ket{1}\rightarrow\ket{0}$. The excitation number
\begin{equation}
N = \sigma_{+} \sigma_{-} + \int \d\omega\, b^\dag_{\omega} b_{\omega}
\end{equation}
is conserved, $[N,H]=0$, thus, the eigenspaces of $N$ with eigenvalues $0,1,2,\dots$ are reducing subspaces for the Hamiltonian $H$ which splits into a direct sum of operators.
The eigenspace corresponding to the eigenvalue $N=0$ is one dimensional and is spanned by the vector $\ket{1,\mathrm{vac}}:=\ket{1}\otimes\ket{\mathrm{vac}}$, while the eigenspace corresponding to the eigenvalue $N=1$ (one-excitation sector) is the linear span of the vectors $\ket{0,\mathrm{vac}}:=\ket{0}\otimes\ket{\mathrm{vac}}$ and $\ket{1,\omega}:=\ket{1}\otimes b^\dag_{\omega} \ket{\mathrm{vac}}$. The latter subspace is isomorphic to $\mathbb{C}\oplus L^2(\omega)$ and the component of $H$ in it is known as the Friedrichs-Lee model~\cite{FriedLee}. Its properties are extensively studied in the references pointed out in the main text. The eigenspaces with higher excitation numbers are spanned by states with at least one photon. For example the two-excitation sector is spanned by $\ket{0}\otimes b^\dag_{\omega} \ket{\mathrm{vac}}$ and $\ket{1}\otimes b^\dag_{\omega} b^\dag_{\omega'}\ket{\mathrm{vac}}$.

We will assume that the initial state of the bath is the vacuum $\ket{\mathrm{vac}}$. Thus we will focus on the reduced dynamics induced by the Hamiltonian~\eqref{eq:ham} on a state $\rho\otimes\ket{\mathrm{vac}}\!\bra{\mathrm{vac}}$, by tracing out the bath. We will evaluate the following quantity:
\begin{equation}\label{eqn:Lambdat}
\rho(t)= \Lambda_t (\rho) = \operatorname{tr}_{\mathrm{bath}}\left(\e^{-\i tH}\rho\otimes\ket{\mathrm{vac}}\!\bra{\mathrm{vac}}\e^{\i tH}\right),
\end{equation}
with $\rho$ being an arbitrary density matrix of the qubit,
\begin{equation}
\rho=\sum_{j=0}^1\rho_{j\ell}\ket{j}\!\bra{\ell}=\begin{pmatrix}
\rho_{00}&\rho_{01}\\
\rho_{10}&\rho_{11}
\end{pmatrix}
\end{equation}
with 
\begin{equation}
\rho=\rho^\ast, \quad \rho \geq 0, \quad \mathrm{tr} (\rho) = \rho_{00}+\rho_{11}=1.
\end{equation}
As such, the evolved density matrix $\rho(t)$ will read
\begin{equation}\label{eqn:rhot}
\rho(t)= \sum_{j,\ell=0}^1 \rho_{j\ell}\,\mathrm{tr}_{\mathrm{bath}}\left(\e^{-\i tH}\ket{j,\mathrm{vac}}\!\bra{\ell,\mathrm{vac}}\e^{\i tH}\right),
\end{equation}
where $\ket{j,\mathrm{vac}}:=\ket{j}\otimes  \ket{\mathrm{vac}}$, for all $j=0,1$. Consequently, we need to compute
\begin{equation}
\e^{-\i tH}\ket{0,\mathrm{vac}},\qquad\e^{-\i tH}\ket{1,\mathrm{vac}}
\end{equation}
for all $t$, i.e. the evolution of $\ket{0,\mathrm{vac}}$ and $\ket{1,\mathrm{vac}}$ under the action of the Hamiltonian $H$. First of all, notice that
\begin{eqnarray}
H\ket{1,\mathrm{vac}}&=&0,\nonumber\\ H\ket{0,\mathrm{vac}}&=&\omega_0\ket{0,\mathrm{vac}}+\int \d\omega\, g(\omega)\ket{1,\omega}, \nonumber\\ H\ket{1,\omega}&=&\omega\ket{1,\omega}+g(\omega)^*\ket{0,\mathrm{vac}},
\end{eqnarray}
therefore the evolution of state $\ket{1,\mathrm{vac}}$ 
is trivial, 
\begin{equation}\label{eq:evol1}
\e^{-\i tH}\ket{1,\mathrm{vac}}=\ket{1,\mathrm{vac}}, 
\end{equation}
while the components $\ket{1,\omega}$ and $\ket{0,\mathrm{vac}}$ evolve nontrivially, without mixing with the previous component. The Schr\"odinger equation
for a global time-dependent state of the form
\begin{equation}
\ket{\Psi(t)}=a(t)\ket{0,\mathrm{vac}}+\int \d\omega\, c(t,\omega)\ket{1,\omega}
\end{equation}
reads
\begin{eqnarray}
& &\i\,\dot{a}(t)\ket{0,\mathrm{vac}}+\i\int \d\omega\, \dot c(t,\omega)\ket{1,\omega}\nonumber\\
& & \qquad\quad=\int \d\omega\, \Big( a(t)\,g(\omega)+\omega\,c(t,\omega)\Big)\ket{1,\omega}
\nonumber\\
& & \qquad \quad \;  +\left(\omega_0\,a(t)+\int \d\omega\, g(\omega)^* c(t,\omega)\right)\ket{0,\mathrm{vac}},
\end{eqnarray}
finally yielding a system of coupled differential equations in $a(t)$ and $c(t,\omega)$:
\begin{equation}
\begin{dcases}
\i\,\dot a(t)=\omega_0\,a(t)+\int \d\omega'\, g(\omega')^* c(t,\omega')\\
\i\,\dot c(t,\omega)=g(\omega) \, a(t) +\omega\,c(t,\omega)
\end{dcases}
\end{equation}
This is exactly the same differential equation that is obtained in~\cite{FriedLee}, albeit in a much more general case, for the generic state of a Friedrichs-Lee Hamiltonian; in this sense, as stated in the main text, our system is a ``variation'' of the Friedrichs-Lee model~\cite{Fried,Lee}. The solution of this system was found explicitly in~\cite{unbounded}. In particular by  choosing as an initial condition the state $\ket{\Psi(0)}=\ket{0,\mathrm{vac}}$, i.e. $a(0)=1$ and $c(0,\omega)=0$, and by taking the Fourier-Laplace transform, for $z\in \mathbb{C}$ with $\operatorname{Im} z>0$,
\begin{eqnarray}
\hat{a}(z) &=& \i \int_0^{+\infty} \d t\, \e^{\i t z} a(t), \nonumber\\ 
\hat{c}(z,\omega) &=& \i \int_0^{+\infty} \d t\, \e^{\i t z} c(t,\omega),
\end{eqnarray}
we get
\begin{equation}
\begin{dcases}
z \hat{a}(z) + 1 =\omega_0\,\hat{a}(z)+\int \d\omega'\, g(\omega')^* \hat{c}(z,\omega');\\
z \hat{c}(z,\omega)=g(\omega) \, \hat{a}(z) +\omega\, \hat{c}(z,\omega).
\end{dcases}
\end{equation}
By plugging the second equation into the first we have
\begin{equation}
\hat{a}(z) = \frac{1}{\omega_0 - z - \Sigma_0(z)},
\end{equation}
where 
\begin{equation}
\Sigma_0(z) = \int \d\omega\, \frac{|g(\omega)|^2}{\omega-z}
\end{equation}
is the bare \emph{self-energy} function. The latter is well defined for $\operatorname{Im}z>0$ as far as 
$\int \d\omega\, |g(\omega)|^2/(|\omega|+1) <+\infty$,
which is the case  if the form factor $g(\omega)$ is a square integrable function. If the form factor $g(\omega)$ is not a square integrable function (e.g. flat form factor $g(\omega)= \mathrm{const}$) the bare self-energy $\Sigma_0(z)$ diverges  and a renormalization procedure is required. More precisely, one should express $\hat{a}(z)$ in terms of dressed quantities $\tilde{\omega}_0$, $\Sigma(z)$, instead of bare ones $\omega_0, \Sigma_0(z)$, namely,
\begin{equation}
\hat{a}(z) = \frac{1}{\tilde{\omega}_0 - z - \Sigma(z)}, 
\end{equation}
where
\begin{equation}
\tilde{\omega}_0 = \omega_0 +\delta\omega_0, \qquad \Sigma(z)  = \Sigma_0(z) + \delta\omega_0,
\label{eq:renorm}
\end{equation}
with $\delta\omega_0$ a suitable renormalization constant. 
By choosing for convenience the subtraction point at $z=\i$, that is  $\delta \omega_0 = \operatorname{Re} \Sigma_0 ( \i)$, the dressed self-energy function is
\begin{eqnarray}\label{eq:self}
\Sigma(z)&=&\int \d\omega\, |g(\omega)|^2 \left(\frac{1}{\omega-z}-\frac{\omega}{\omega^2+1}\right) 
\nonumber\\&=&\int \d\omega\, |g(\omega)|^2 \frac{1 + \omega z}{(\omega-z)(\omega^2+1)}.
\end{eqnarray}
Notice that the dressed self-energy $\Sigma(z)$ in~\eqref{eq:self} is well defined even for a flat form factor $g(\omega)= \mathrm{const}$. In this case  the bare quantities $\omega_0$ and $\Sigma_0(z)$, as well as the energy shift $\delta\omega_0$, diverge  but the sums  in~\eqref{eq:renorm} are finite and give a well-defined model characterized by $\Sigma(z)$ (and hence $g(\omega)$) and by the dressed qubit energy $\varepsilon_0$.  

By transforming back to the time domain one finally gets
\begin{equation}
\label{eq:at}
a(t)=\frac{1}{2\pi\i}\int_{\mathbb{R}+\i y}\frac{\e^{-\i zt}}{\tilde{\omega}_0-z-\Sigma(z)}\,\mathrm{d}z,
\end{equation}
with an arbitrary $y>0$. The above heuristic derivation can be made fully rigorous~\cite{FriedLee} and one can show that for every $\tilde{\omega}_0\in \mathbb{R}$ and for every form factor $g(\omega)$ satisfying the growth condition 
\begin{equation}
\int \d\omega\, \frac{|g(\omega)|^2}{\omega^2+1} <+\infty,
\end{equation}
the Friedrichs-Lee Hamiltonian is self-adjoint and thus yields a unitary evolution with a survival amplitude given by~\eqref{eq:at}. For the sake of a simple notation, in the following we will denote the dressed qubit energy by $\omega_0$. 

Therefore
\begin{equation}\label{eq:evol2}
\e^{-\i tH}\ket{0,\mathrm{vac}}=a(t)\ket{0,\mathrm{vac}}+\int  \d\omega\, c(t,\omega)\ket{1,\omega} ,
\end{equation}
with $a(t)$ given by~\eqref{eq:at}. Notice that, since the global evolution is unitary,
\begin{equation}
\int|c(t,\omega)|^2\,\mathrm{d}\omega=1-|a(t)|^2.
\label{eq:unitr}
\end{equation}
Having evaluated both $\e^{-\i tH}\ket{0,\mathrm{vac}}$ and $\e^{-\i tH}\ket{1,\mathrm{vac}}$ (see Eqs.~\eqref{eq:evol1} and~\eqref{eq:evol2}), we have
\begin{widetext}
\begin{eqnarray}
\e^{-\i tH}\ket{0,\mathrm{vac}}\!\bra{0,\mathrm{vac}}\e^{\i tH}&=&|a(t)|^2\ket{0,\mathrm{vac}}\!\bra{0,\mathrm{vac}}+\iint \d \omega\,\d\omega' \,  c(t,\omega) c(t,\omega')^* \ket{1,\omega}\!\bra{1,\omega'}\nonumber\\
& & + a(t)\int \d\omega\, c(t,\omega)^* \ket{0,\mathrm{vac}}\!\bra{1,\omega}+a(t)^*\int \d\omega\, c(t,\omega) \ket{1,\omega}\!\bra{0,\mathrm{vac}};\\
\e^{-\i tH}\ket{1,\mathrm{vac}}\!\bra{1,\mathrm{vac}}\e^{\i tH}&=&\ket{1,\mathrm{vac}}\!\bra{1,\mathrm{vac}};\\
\e^{-\i tH}\ket{0,\mathrm{vac}}\!\bra{1,\mathrm{vac}}\e^{\i tH}&=&a(t)\ket{0,\mathrm{vac}}\!\bra{1,\mathrm{vac}}+\int  \d\omega\, c(t,\omega)\ket{1,\omega}\!\bra{0,\mathrm{vac}}.
\end{eqnarray}
By tracing out the bath we get
\begin{eqnarray}
\operatorname{tr}_{\mathrm{bath}} (\e^{-\i tH}\ket{0,\mathrm{vac}}\!\bra{0,\mathrm{vac}}\e^{\i tH})&=&|a(t)|^2\ket{0}\!\bra{0}+\int \d \omega\, |c(t,\omega)|^2  \ket{1}\!\bra{1};\\
\operatorname{tr}_{\mathrm{bath}} (\e^{-\i tH}\ket{1,\mathrm{vac}}\!\bra{1,\mathrm{vac}}\e^{\i tH})&=&\ket{1}\!\bra{1};\\
\operatorname{tr}_{\mathrm{bath}} (\e^{-\i tH}\ket{0,\mathrm{vac}}\!\bra{1,\mathrm{vac}}\e^{\i tH})&=&a(t)\ket{0}\!\bra{1},
\end{eqnarray}
\end{widetext}
and recalling~\eqref{eqn:rhot} and~\eqref{eq:unitr}
we finally get
\begin{equation}
\rho(t)= \Lambda_t (\rho) =
\begin{pmatrix}
|a(t)|^2\rho_{00}& a(t)  \rho_{01}\\
a(t)^*\rho_{10}&\rho_{11}+(1-|a(t)|^2)\rho_{00}
\end{pmatrix}.
\end{equation}
Now we define two real functions $\gamma(t)$ and $\varepsilon(t)$ such that $a(t)$ can be rewritten as
\begin{equation}
a(t) = \exp\bigg(-\int_0^t \d s\, \Big(\frac{\gamma(s) }{2}  + \i \varepsilon(s) \Big) \bigg), 
\end{equation}
or equivalently such that
\begin{equation}
\frac{\dot a(t)}{a(t)} = - \frac{\gamma(t) }{2}  - \i \varepsilon(t).
\end{equation}
By a simple computation one gets
\begin{equation}
\gamma(t) = -\frac{2}{|a(t)|} \frac{\d}{\d t} |a(t)|,  \quad \varepsilon(t)= \frac{\i}{\operatorname{sgn} (a(t))} \frac{\d}{\d t} \operatorname{sgn} (a(t))
\end{equation}
with $\operatorname{sgn} (z) = z/|z|$. Using these functions the derivative of $\rho(t)$ reads
\begin{equation}
\dot\rho(t)=\begin{pmatrix}
-\gamma(t) \rho_{00}(t)& \big(-\frac{\gamma(t) }{2}  - \i \varepsilon(t) \big)\rho_{01}(t)\\
\big(-\frac{\gamma(t) }{2}  + \i \varepsilon(t) \big) \rho_{10}(t) & \gamma(t) \rho_{00}(t)
\end{pmatrix},
\end{equation}
where $\rho_{00}(t):=|a(t)|^2\rho_{00}$, $\rho_{01}(t):=a(t)  \rho_{01}$, $\rho_{10}(t):=a(t)^*  \rho_{10}$ and $\rho_{11}(t):=\rho_{11}+(1-|a(t)|^2)\rho_{00}$.
Therefore $\dot\rho(t)$ can be written as
\begin{equation}
\dot\rho(t) = -\i \varepsilon(t) \bigl[H_{\mathrm{q}}, \rho(t) \bigr]  - \gamma(t) \mathcal{L} (\rho(t)),
\end{equation}
with
\begin{equation}
H_{\mathrm{q}} = 
\ket{0}\!\bra{0} , 
\end{equation}
and
\begin{equation}
\mathcal{L}(\rho) =  - \sigma_{-} \rho \sigma_{+} + \frac{1}{2} \bigl\{ \sigma_{+} \sigma_{-}, \rho \bigr\}, 
\end{equation}
where, as usual, the square brackets denote the commutator while the curly brackets denote the anticommutator.
Therefore the quantum channel $\Lambda_t$ in~\eqref{eqn:Lambdat} describing the evolution of the qubit has a generator in the GKLS form,
\begin{equation}
-\i \varepsilon(t) \operatorname{ad}_{H_{\mathrm{q}}}  -\gamma(t) \mathcal{L},
\end{equation}
with time dependent coefficients $\varepsilon(t)$ and $\gamma(t)$, where $\operatorname{ad}_{H_{\mathrm{q}}}(\rho)=\bigl[H_{\mathrm{q}}, \rho \bigr]$.
Since $\operatorname{ad}_{H_{\mathrm{q}}}$ and $\mathcal{L}$ commute, we have
\begin{eqnarray}
\Lambda_t &=& \exp\bigg(-\int_0^t \d s\, \Big(\gamma(s) \mathcal{L}  + \i \varepsilon(s)  \operatorname{ad}_{H_{\mathrm{q}}} \Big) \bigg) =\nonumber\\&=& \exp\bigg(\ln (|a(t)|^2) \mathcal{L}  + \i \arg (a(t))  \operatorname{ad}_{H_{\mathrm{q}}} \bigg) .
\end{eqnarray}
In the case $\gamma(t)=\gamma_0 = \mathrm{const}$ and $\varepsilon(t)=\varepsilon_0 = \mathrm{const}$, i.e. $a(t)=\e^{-(\gamma_0/2 +\i\varepsilon_0)t}$, we have
\begin{equation}
\Lambda_t =\e^{- t( \gamma_0 \mathcal{L} + \i\varepsilon_0\operatorname{ad}_{H_{\mathrm{q}}}) }
\end{equation}
and the semigroup property, i.e. $\Lambda_t \Lambda_s=\Lambda_{t+s}$ for all $t,s \geq 0$, is satisfied and hence the channel is Markovian; this is the  amplitude-damping channel. More generally, the semigroup property would be satisfied if and only if  $a(t+s)=a(t)a(s)$  for all $t,s \geq 0$, which is not satisfied in general, thus preventing the channel to be Markovian.

\subsection{Coupling and evolution}
Eq.~\eqref{eq:at} implies that the value of $a(t)$ is ultimately determined by the self-energy $\Sigma(z)$, which in turn depends on the square modulus $|g(\omega)|^2$ of the form factor via Eq.~\eqref{eq:self}. In fact, the correspondence between $\Sigma(z)$ and $|g(\omega)|^2$ is unique, as discussed in~\cite{FriedLee} and references therein: $|g(\omega)|^2$ can be reconstructed from the self-energy via
\begin{equation}\label{eq:gself}
|g(\omega)|^2=\frac{1}{\pi}\lim_{\delta \downarrow 0}\operatorname{Im}\,\Sigma(\omega+\i \delta).
\end{equation}
As a first example, by setting $g(\omega)=\sqrt{\gamma_0/2\pi}$ for some $\gamma_0>0$, we have $\Sigma(z)=\i\frac{\gamma_0}{2}$ whenever $\mathrm{Im}\,z>0$ and thus, substituting in Eq.~\eqref{eq:at}, one immediately obtains
\begin{equation}
a(t)=\e^{-\left(\frac{\gamma_0}{2}+\i\varepsilon_0\right)t},
\end{equation}
i.e. a flat coupling yields an exponential decay of the damping rate $a(t)$ at all times.

Let us examine the case of a periodic coupling, written in a Fourier cosine series as
\begin{equation}
|g(\omega)|^2=\frac{\gamma_0}{2\pi}\left(1+2\sum_{n=1}^\infty c_n\cos nT\omega\right)
\end{equation}
for some family of real coefficients $\{c_n\}_{n=1}^\infty$ chosen in such a way that the series is absolutely convergent and positive for all $\omega$. The corresponding self-energy reads
\begin{equation}
\Sigma(z)=\frac{\i\gamma_0}{2}\left(1+2\sum_{n=1}^\infty c_n\e^{\i nTz}\right),
\end{equation}
which can be verified immediately by Eq.~\eqref{eq:gself}. With this choice of self-energy, from
\begin{equation}
\hat{a}(z)=\frac{1}{\varepsilon_0-z-\Sigma(z)},
\end{equation}
by a simple calculation we get
\begin{equation}
\hat{a}(z)=\frac{1}{\varepsilon_0-z-\frac{\i\gamma_0}{2}}+\i\gamma_0\sum_{n=1}^\infty c_n\e^{\i nTz}\frac{\hat{a}(z)}{\varepsilon_0-z-\frac{\i\gamma_0}{2}},
\end{equation}
which implies
\begin{eqnarray}\label{eq:recursion}
a(t)&=&\e^{-\left(\frac{\gamma_0}{2}+\i\varepsilon_0\right)t}
\nonumber\\& & -\gamma\sum_{n=1}^\infty c_n\,\theta(t-nT)\left[a\star e^{-\left(\frac{\gamma_0}{2}+\i\varepsilon_0\right)\cdot}\right](t-nT),\nonumber\\
\end{eqnarray}
where $\theta(t)$ is the Heaviside step function, and $\star$ is the convolution product evaluated at $t-nT$. From this equation it is already clear that $a(t)$ will be exactly exponential up to $t=T$, thereafter non-exponential corrections will add up.

The solution of this equation can be found by means of a proper ansatz:
\begin{eqnarray}\label{eq:ansatz}
a(t)&=&\e^{-\left(\i\varepsilon_0+\frac{\gamma_0}{2}\right)t}
\nonumber\\
& & +\sum_{n=1}^\infty \e^{-\left(\i\varepsilon_0+\frac{\gamma_0}{2}\right)(t-nT)}\theta(t-nT)\phi_n(\gamma_0(t-nT)),\nonumber\\
\end{eqnarray}
where $\phi_n(x)$ is some function to be evaluated. By imposing Eq.~\eqref{eq:recursion} for the function in~\eqref{eq:ansatz}, one obtains a solvable recursion equation in $n$ for the functions $\phi_n(x)$ which finally yields
\begin{equation}
\label{eq:phin}
\phi_n(x)=\sum_{m=1}^nb_n^{(m)}\frac{(-x)^m}{m!},
\end{equation}
where the coefficients $b_n^{(m)}$ for $m=1,\dots,n$, are
\begin{equation}
\label{eq:b}
b_n^{(m)}=\sum_{(h_1,\dots,h_m)\in I_n^m}\left(\prod_{i=1}^mc_{h_i}\right),
\end{equation}
with $I_n^m$ being the set of all ordered $m$-tuples of strictly positive integers that sum to $n$, i.e. 
\begin{equation}
I_n^m=\{(h_1,\dots,h_m)\in\mathbb{N}^m \setminus\{0\}:   h_1+\dots +h_m=n \},
\end{equation}
that is, the positive integer elements of the $m$-dimensional simplex with edge length $n$. Notice that the cardinality of this set is
\begin{equation}
\#\left(I_n^m\right)=\binom{n-1}{m-1},
\label{eq:card}
\end{equation}
as can be proven through the usual stars-and-bars argument. By these formulas, we are finally able to compute the polynomials $\phi_n(x)$ for the two examples in the main text. 

\paragraph{Single nonzero coefficient (sinusoidal measure)} In the case 
\begin{equation}
c_1=-\frac{\alpha}{2},\qquad c_{n}=0 \quad \forall n\geq2,
\end{equation}
for some  $|\alpha| \leq 1$, we have only one nonzero coefficient, therefore the only elements which must be taken into account in the sum are $m$-tuples in the form $(1,1,\dots,1)$, which do belong to the simplex $I_n^m$ if and only if $n=m$; as a result, the only nonzero coefficients $b_n^{(m)}$ are those with $n=m$, with
\begin{equation}
b_n^{(n)}=\prod_{i=1}^nc_{1,\alpha}=\frac{(-\alpha)^n}{2^n},
\end{equation}
hence
\begin{equation}
\phi_n(x)=\frac{1}{n!}\left(\frac{\alpha x}{2}\right)^n.
\end{equation}

\paragraph{Exponentially decaying coefficients (smoothed Dirac measure)}
In the case
\begin{equation}
c_n=\e^{-\beta n} \quad \forall n \in \mathbb{N},
\end{equation}
for some $\beta \geq0$, the coefficients satisfy the property
\begin{equation}
\prod_{i=1}^m c_{h_i}=c_{h_1+h_2+\dots+h_m}
\end{equation}
and hence, by Eqs.~\eqref{eq:b} and~\eqref{eq:card},
\begin{equation}
b_n^{(m)}=\binom{n-1}{m-1}\e^{-\beta n},
\end{equation}
thus implying
\begin{equation}
\phi_n(x)=\e^{-\beta n}\sum_{m=1}^n\binom{n-1}{m-1}\frac{(-x)^m}{m!};
\end{equation}
this implies that the non-markovian contributions to the survival amplitude have the same functional expression for all $\beta$, up to a total weight $\e^{-\beta n}$ which suppresses such contributions as $n$ grows, provided that $\beta>0$; as a result, the larger $\beta$, the quicker such contributions "switch off", whereas for small $\beta$ those contributions are non-negligible for a longer time. In particular, in the limit $\beta\to\infty$ all non-markovian contributions vanish and we recover the exponential decay at all times.


\begin{thebibliography}{99}

\bibitem{Modi1}  G. A. L. White, C. D. Hill, F. A. Pollock, L. C. L. Hollenberg, K. Modi, \textit{Experimental non-Markovian process characterisation and control on a quantum processor}, 	arXiv:2004.14018 [quant-ph].

\bibitem{LiLi}Li Li, M. J. W. Hall, H. M. Wiseman, \textit{Concepts of quantum non-Markovianity: a hierarchy}, Physics Reports, \textbf{759}, 1 (2018).

\bibitem{Dorner02}
U. Dorner, P. Zoller,
\textit{Laser-driven atoms in half-cavities}, 
Phys. Rev A \textbf{66}, 023816 (2002).

\bibitem{zeta}
N. L\"orch, F.V. Pepe, H. Lignier, D. Ciampini, R. Mannella, O. Morsch, E. Arimondo, P. Facchi, G. Florio, S. Pascazio, S. Wimberger,
\textit{Wave-function-renormalization effects in resonantly enhanced tunneling},
Phys. Rev. A \textbf{85}, 053602 (2012) 

\bibitem{Tufarelli14}
T. Tufarelli, M. S. Kim, F. Ciccarello,
\textit{Non-Markovianity of a quantum emitter in front of a mirror}, 
Phys. Rev. A \textbf{90}, 012113 (2014)

\bibitem{qdot_review}
P. Lodahl, S. Mahmoodian, and S. Stobbe, \textit{Interfacing single photons and single quantum dots with photonic nanostructures}, Rev. Mod. Phys. \textbf{87}, 347 (2015).

\bibitem{kimble1}
J. S. Douglas, H. Habibian, C.-L. Hung, A. V. Gorshkov, H. J. Kimble, D. E. Chang, \textit{Quantum many-body models with cold atoms coupled to photonic crystals}, Nat. Photonics \textbf{9}, 326 (2015).

\bibitem{ck}
A. Gonz\'alez-Tudela, V. Paulisch, H. J. Kimble, and J. I. Cirac, \textit{Efficient Multiphoton Generation in Waveguide Quantum Electrodynamics}, Phys. Rev. Lett. \textbf{118}, 213601 (2017).

\bibitem{waveguide}
P. Facchi, M.S. Kim, S. Pascazio, F.V. Pepe, D. Pomarico, T. Tufarelli,
\textit{Bound states and entanglement generation in waveguide quantum electrodynamics},
Phys. Rev. A \textbf{94}, 043839 (2016) 

\bibitem{Fang18}
Y.-L. L Fang, F. Ciccarello, H. U. Baranger,
\textit{Non-Markovian dynamics of a qubit due to single-photon scattering
	in a waveguide}, 
New J. Phys. \textbf{20},  043035 (2018).

\bibitem{n-wave}
P. Facchi, D. Lonigro, S. Pascazio, F.V. Pepe, D. Pomarico,
\textit{Bound states in the continuum for an array of quantum emitters},
Phys. Rev. A \textbf{100}, 023834 (2019).

\bibitem{Liu2018}
Z.-D. Liu, H. Lyyra, Y.-N. Sun, B.-H. Liu, C.-F. Li, G.-C. Guo, S. Maniscalco, J. Piilo,
\textit{Experimental implementation of fully controlled dephasing dynamics and synthetic spectral densities},
Nature Comm. \textbf{9}, 3453 (2018).

\bibitem{Liu2011}
B.-H. Liu, L. Li, Y.-F. Huang, C.-F. Li, G.-C. Guo, E.-M. Laine, H.-P. Breuer, J. Piilo, 
\textit{Experimental control of the transition from Markovian to non-Markovian dynamics of open quantum systems}, Nature Phys. \textbf{7}, 931 (2011).

\bibitem{Orieux2015}
A. Orieux, A. D'Arrigo, G. Ferranti, R. Lo Franco, G. Benenti, E. Paladino, G. Falci, F. Sciarrino, P. Mataloni,
\textit{Experimental on-demand recovery of entanglement by local operations within non-Markovian dynamics}, Sci. Rep. \textbf{5}, 8575 (2015).

\bibitem{Bernardes2015}
N. K. Bernardes, A. Cuevas, A. Orieux, C. H. Monken, P. Mataloni, F. Sciarrino, M. F. Santos, 
\textit{Experimental observation of weak non-Markovianity},
Sci. Rep. \textbf{5}, 17520 (2015).

\bibitem{Cuevas19}
A. Cuevas, A. Geraldi, C. Liorni, L. D. Bonavena, A. De Pasquale, F. Sciarrino, V. Giovannetti, P. Mataloni,
\textit{All-optical implementation of collision-based evolutions of open quantum systems} 
Sci. Rep. \textbf{9}, 3205 (2019).

\bibitem{Zou2013}
C.-L. Zou, X.-D. Chen, X. Xiong, F.-W. Sun, X.-B. Zou, Z.-F. Han, and G.-C. Guo,
\textit{Photonic simulation of system-environment interaction: Non-Markovian processes and dynamical decoupling}, 
Phys. Rev. A \textbf{88}, 063806 (2013).

\bibitem{ladder}
A. Crespi, F.V. Pepe, P. Facchi, F. Sciarrino, P. Mataloni, H. Nakazato, S. Pascazio, R. Osellame,
\textit{Experimental investigation of quantum decay at short, intermediate and long times via integrated photonics},
Phys. Rev. Lett. \textbf{122}, 130401 (2019). 

\bibitem{GKLS}
V. Gorini, A. Kossakowski, E. C. G. Sudarshan, 
\textit{Completely positive dynamical semigroups of N-level systems},
J. Math. Phys. \textbf{17}, 821 (1976).

\bibitem{Lindblad}
G. Lindblad, 
\textit{On the generators of quantum dynamical semigroups},
Commun. Math. Phys. \textbf{48}, 119 (1976).

\bibitem{Modi}S. Milz, M. S. Kim, F. A. Pollock, and K. Modi, \textit{Completely Positive Divisibility Does Not Mean Markovianity}, Phys. Rev. Lett. \textbf{123}, 040401 (2019).

\bibitem{Wolf}
M. M. Wolf, J. Eisert, T. S. Cubitt, J. I. Cirac, 
\textit{Assessing Non-Markovian Quantum Dynamics},
Phys. Rev. Lett. \textbf{101}, 150402 (2008).

\bibitem{Wolf2} 
T. S. Cubitt, J. Eisert, M. M. Wolf, \emph{The Complexity of Relating Quantum Channels to Master Equations}, Communications in Mathematical Physics \textbf{310}, 383 (2012).

\bibitem{Precursor} S. Campbell, M. Popovic, D. Tamascelli and B. Vacchini, \textit{Precursors of non-Markovianity}, New J. Phys. \textbf{21} 053036 (2019).

\bibitem{FriedLeeProc}
D. Lonigro, P. Facchi, M. Ligab\`{o}, 
\textit{The Friedrichs-Lee model and its singular coupling limit},
Proceedings \textbf{12}(1), 17 (2019).

\bibitem{FriedLee}
P. Facchi, M. Ligab\`{o} and D. Lonigro, 
\textit{Spectral properties of the singular Friedrichs-Lee Hamiltonian},
arXiv:1910.05957 [math-ph] (2019).

\bibitem{unbounded}
C. Arenz, D. Burgarth, P. Facchi, R. Hillier,
\textit{Dynamical decoupling of unbounded Hamiltonians},
J. Math. Phys. \textbf{59}, 032203 (2018).

\bibitem{Milonni83} 
P. W. Milonni, J. R. Ackerhalt, H. W. Galbraith, Mei-Li Shih,
\textit{Exponential decay, recurrences, and quantum-mechanical spreading in a quasicontinuum model},
Phys. Rev. A \textbf{28}, 32 (1984).

\bibitem{Fried}
K. Friedrichs, 
\textit{On the perturbation of continuous spectra},
Commun. Pur. Appl. Math. \textbf{1}, 361 (1948).

\bibitem{Lee}
T. Lee, 
\textit{Some Special Examples in Renormalizable Field Theory},
Phys. Rev. \textbf{95}, 1329 (1954).



\end{thebibliography}
\end{document}